# Efficient Multi-Year Security Constrained AC Transmission Network Expansion Planning

Soumya Das, *Student Member, IEEE*, Ashu Verma, *Sr. Member, IEEE* and P. R. Bijwe, *Sr. Member, IEEE*

*Abstract*—Solution of multi-year, dynamic AC Transmission network expansion planning (TNEP) problem is gradually taking center stage of planning research owing to its potential accuracy. However, computational burden for a security constrained AC TNEP is huge compared to that with DC TNEP. For a dynamic, security constrained AC TNEP problem, the computational burden becomes so very excessive that solution for even moderately sized systems becomes almost impossible. Hence, this paper presents an efficient, four-stage solution methodology for multi-year, network N-1 contingency and voltage stability constrained, dynamic ACTNEP problems. Several intelligent logical strategies are developed and applied to reduce the computational burden of optimization algorithms. The proposed methodology is applied to Garver 6, IEEE 24 and 118 bus systems to demonstrate its efficiency and ability to solve TNEP for varying system sizes.

*Index Terms*—Power system expansion planning, multi-year planning, intelligent algorithms, network security constraints.

## I. Nomenclature

*A. Variables, Sets and Parameters Related to TNEP*

$v_d$ — Total investment cost referred to first year of planning
$T_y$ — Total number of years in planning horizon
$y$ — Year of planning
$k$ — Contingency state, $k = 0$ denotes base case

$\forall\ year,\ y$:

$C_d^y$ — Discount factor for investment cost
$v^y$ — Total investment cost
$v_{lc}^y$ — Line addition cost
$l^y$ — Power corridor between two buses
$\Omega$ — Set of all power corridors
$C_l^y$ — Cost of line addition in the $l^{th}$ power corridor
$n_l^y$ — Number of additional lines in the $l^{th}$ power corridor
$N_{ld}$ — Set of load buses of the system
$P_{Dm}^y$ — Vector of real power demand
$Q_{Dm}^y$ — Vector of reactive power demand
$q_{rc}^y$ — Vector of capacity of additional reactive sources
$L^y$ — Base case L-index value of the system

$\forall\ year,\ y$ and $\forall contingency,\ k$:

$P_{in_k}^y$ — Vector of real power injections
$P_{Gn_k}^y$ — Vector of real power generations
$Q_{in_k}^y$ — Vector of reactive power injections
$Q_{Gn_k}^y$ — Vector of reactive power generations
$V_k^y$ — Vector of bus voltage magnitudes
$\theta_k^y$ — Vector of bus voltage angles
$n_k^y$ — Vector of system circuits
$S_{ly}^{kfr}$ — Sending end MVA flow in each line of $l^{th}$ corridor
$S_{ly}^{kto}$ — Receiving end MVA flow in each line of $l^{th}$ corridor
$\bar{n}_l$ — Maximum limit on new lines in $l^{th}$ power corridor
$\omega_{cont}$ — Total number of contingencies in a system
$N_{bus}$ — Set of all buses in a system

*B. Parameters Related to Metaheuristic Algorithm*

$cs_N$ — Population/colony size
$E_h$ — Number of neighbours
$iter$ — Maximum number of iterations per trial
$lim$ — Parameter for generation of scout bees
$w_g$ — Parameter to control the effect of best solution
$tp$ — Approximate time of solution per trial

## II. Introduction

POWER system networks worldwide have encountered fundamental changes owing to their unbundling and deregulation. In addition, environmental concerns to reduce greenhouse gas emissions are encouraging increasing the amount of integration of renewable sources after the enactment of Kyoto protocol [1]. Solution of multi-year, dynamic transmission network expansion planning (TNEP) can adequately address this by providing the network planners not only the information on which lines to construct, but also the time of its construction within the planning horizon so that the overall investment cost is minimized and the uncertainties in future load/generation can be addressed.

However, owing to its NP-hard, mixed-integer, combinatorial nature, solution to such an optimization problem is extremely complex. Consideration of network contingencies make it even formidable to solve. Extensive research has been conducted on this topic in past few decades, although, with several simplifications. Therefore, in existing literatures, researchers have mostly used approximated DC formulation to solve the problem within a manageable time frame. A detailed and comprehensive review of the existing TNEP literature is presented in [2]. Metaheuristic methods are generally applied [3] – [8]. Mixed integer linear programming (MILP) approach with DC model has also been used by several researchers to solve multi-year TNEP [9] – [14]. Also, dynamic generation and transmission expansion planning in a DC framework are

Soumya Das is currently a Ph. D scholar in the Centre for Energy Studies, IIT Delhi, India (email: soumya.das@ces.iitd.ac.in)
Ashu Verma is with the Centre for Energy Studies, IIT Delhi, India (e-mail: averma@ces.iitd.ac.in).
P. R. Bijwe is with the Department of Electrical Engineering, IIT Delhi, India (e-mail: prbijwe@ee.iitd.ac.in).

solved in [15] – [17].

Due to increased computational burden experienced in solving multi-year TNEP even in DC framework, several strategies for effective reduction of the same have also been explored by researchers. Heuristic strategies to solve the problem is presented by authors in [18]. In [19], several strategies to reduce the combinatorial search space is proposed in solving multi-year TNEP; and, a two-stage solution methodology for solving similar problem is presented in [20].

Primary drawback of using DCTNEP plans is that, as planning is obtained by neglecting system reactive power flows, direct application of them to AC networks can cause line overloads, and, in the worst case, in absence of proper reactive support, may lead to system voltage collapse. As in current deregulated scenario, maintaining an acceptable system voltage profile is a strict requirement, solution of ACTNEP is gradually gaining interest. However, due to extreme computational burden in solving non-linear network equations, only a few recent literatures explore linearized solution of these problems.

MILP approach to solve security constrained multi-year ACTNEP is explored in [21]. Here, an initial DC solution is reinforced to obtain security constrained AC results. Such a formulation, although very effective in reducing computational burden, leads to sub-optimal solutions. In [22], the author uses mixed integer conic programming (MICP), to solve static and multi-year ACTNEP, without consideration of network contingencies. Solution for a smaller system is obtained very quickly, however, solution time for a medium sized system is extremely large. Therefore, the author recommends the readers to explore different techniques which reduce the overall search space. In [23], MILP is used to solve a similar problem. Main drawback faced here is dealing with high dimensionality and computational burden. Also, in [24], a multi-objective dynamic ACTNEP is solved without consideration of security constraints by multi-objective evolutionary PSO (MEPSO) algorithm. Security constrained, multi-year linearized ACTNEP is solved by MILP in [25]. However, here, the authors have considered a set of only a few selective contingencies instead of all available contingencies for contingency studies.

From the above literature review it is clear that, most of the existing literature focus on solving multi-year TNEP with DC formulation. Due to huge computational burden in solving even DC problems, several strategies have been proposed. Multi-year dynamic ACTNEP is a relatively new area of research, where most of the works consider approximated linearized solution processes to tackle the issue of extreme complexity. Therefore, these are approximate plans, and their direct application to practical systems may cause unintentional security issues. It can be observed from literature that in solving multi-year ACTNEP with security constraints, full non-linear formulation has not been attempted in the past to that extent. This paper addresses the research gap by proposing a four-stage solution methodology which uses several intelligent strategies, to effectively reduce the overall computational burden in solving such problems.

The proposed methodology depends on the fact that, the power corridors which will have new lines, and the total number of new lines in the final AC contingency constrained planning, can be very effectively estimated from: a) Contingency constrained DCTNEP and b) ACTNEP without considering contingencies. Therefore, the first three stages involve the respective solutions of: 1) Base case DCTNEP, 2) Base case ACTNEP, and, 3) Contingency constrained DCTNEP. Intelligent strategies are then formed on the basis of these results to solve the fourth and final stage of security constrained ACTNEP. To demonstrate the potential of the developed framework, computational burden of a metaheuristic, modified artificial bee colony algorithm (MABC) [26], is compared with/without the proposed strategies. However, the strategies are generic enough and can be applied to any other metaheuristic algorithm as well.

Primary contributions of this paper can be summarized as:
1) Development of a novel four-stage solution methodology for efficient solution of security constrained multi-year dynamic ACTNEP.
2) Demonstrate the applicability of the proposed methodology for solution of dynamic ACTNEP problems for different sized systems with drastic reduction of computational burden compared to conventional methods.

Rest of this paper is organized as follows: Section III provides the mathematical modeling, followed by a description of the solution methodology used, in Section IV. Section V describes the proposed algorithm. Section VI provides detailed results of the test systems and relevant discussions. Finally, conclusions and future work are stated in Section VII.

### III. MATHEMATICAL MODELING

In this paper, a multi-year dynamic ACTNEP problem is solved, with approximate reactive power planning (RPP) being considered from the solution of static, sequential ACTNEP. This is because, experience has shown that, the RPP of the final dynamic ACTNEP is not much different compared to the RPP obtained from sequential planning. Such a consideration of approximate RPP reduces the complexity and computational burden in solving dynamic ACTNEP as will be discussed in detail in Section V. The objective thus becomes minimization of the total cost of line additions over the planning horizon [27]:

Minimize:
$$v_d = \sum_{y=1}^{T_y} (C_d^y \times v_{lc}^y) \qquad (1)$$

where, $\forall\, y$,
$$v_{lc}^y = \sum_{l^y \in \Omega} (C_l^y \times n_l^y) \qquad (2)$$

In multi-year TNEP, cost of construction is referred to the first year with appropriate discount factors to account for cost depreciation. Therefore, the objective function for minimization is represented by (1), which is the sum of total investment cost per year multiplied by the respective discount factors, $C_d^y$. Cost of investment in each year, is the line investment cost, $v_{lc}^y$ represented by (2). $C_l^y$ represents the cost of line addition in the $l^{th}$ power corridor in $y^{th}$ year.

Several constraints are required to be satisfied for each year $y$ and for each contingency state $k$. The constraints that govern the above minimization problem can be grouped as follows:

### A. Operational constraints:

These are network power balance constraints at all the buses,
$$\boldsymbol{P_{in}(V,\theta,n)}_k^y - \boldsymbol{P_{Gn_k}^y} + \boldsymbol{P_{Dm}^y} = 0 \tag{3}$$
$$\boldsymbol{Q_{in}(V,\theta,n)}_k^y - \boldsymbol{Q_{Gn_k}^y} + \boldsymbol{Q_{Dm}^y} - \boldsymbol{q_{rc}^y} = 0 \tag{4}$$

In addition, network voltage profile is required to be maintained within a specified upper and lower bound,
$$\boldsymbol{V_{Min}^y} \leq \boldsymbol{V_k^y} \leq \boldsymbol{V_{Max}^y} \tag{5}$$

Further, as voltage stability of a system is a major concern in current deregulated scenarios, a good planning should provide adequate margin of the same. System L-index [28] value provides a fair estimate of its voltage stability. Ranging from 0 to 1, L-index value of 1 indicates system voltage collapse, whereas 0 indicate a very stable system.

However, as L-index is highly nonlinear, accurate realization of network MW voltage stability margin is not possible only through its value. For obtaining an accurate voltage stability margin, proper RPP is required to be solved along with ACTNEP. In this work, a simplistic RPP is solved only to ensure system convergence and adequate reactive support. After an initial investment plan is obtained by the method proposed in this paper, a user can perform a proper RPP with consideration of an accurate voltage stability margin to obtain a final plan. Such a decomposed approach to the problem is essential for managing the computational burden involved in solving dynamic ACTNEP.

Therefore, with a simplistic RPP as is done here, to provide at least an approximate estimation of a good, voltage-stable system, a bound is set on its L-index value. If L-index value of a system is maintained within a low maximum bound (typically 0.4), it can result in an adequately voltage-stable system. Although the boundary value considered is not optimal, the model proposed is general enough, and users can define an optimal limit on L-index values according to their choice. It is enforced only for base case network, as similar enforcement even in the contingency cases may result in a significantly increased investment cost. Further, limiting base case L-index value of a system within a low bound obviously increases voltage stability margin even for contingency cases. Thus,
$$L_{Min}^y \leq L^y \leq L_{Max}^y \tag{6}$$

### B. Equipment constraints:

Equipment constraints include real and reactive power generation limits of the generators and line power flow limits. The generator limits are provided by (7) and (8).
$$\boldsymbol{P_{Gn_{Min}}^y} \leq \boldsymbol{P_{Gn_k}^y} \leq \boldsymbol{P_{Gn_{Max}}^y} \tag{7}$$
$$\boldsymbol{Q_{Gn_{Min}}^y} \leq \boldsymbol{Q_{Gn_k}^y} \leq \boldsymbol{Q_{Gn_{Max}}^y} \tag{8}$$

Line power flow limits are considered as follows:
$\forall l^y \in \Omega \; ; l^y \neq k$,
$$(n_0^y + n_l^y) S_{l^y}^{kfr} \leq (n_0^y + n_l^y) S_{l^y_{Max}} \tag{9}$$
$$(n_0^y + n_l^y) S_{l^y}^{kto} \leq (n_0^y + n_l^y) S_{l^y_{Max}} \tag{10}$$
for $l^y = k, k \neq 0$,
$$(n_0^y + n_l^y - 1) S_{l^y}^{kfr} \leq (n_0^y + n_l^y - 1) S_{l^y_{Max}} \tag{11}$$
$$(n_0^y + n_l^y - 1) S_{l^y}^{kto} \leq (n_0^y + n_l^y - 1) S_{l^y_{Max}} \tag{12}$$

Here, each power corridor is modelled to have lines with exactly similar characteristics. Whenever a line of a different characteristic is added in an existing power corridor, the added line constitutes a separate sub-corridor in between the same buses of the system. Such consideration increases system reliability as it allows the model to track contingencies of different line types in a physical power corridor while performing N-1 contingency analysis.

### C. Physical constraints:

These constraints include physical limitations in a network planning, such as, limits on maximum number of new lines per corridor:
$$\forall l^y \in \Omega, \quad 0 \leq \sum_{y=1}^{T_y} n_l^y \leq \bar{n}_l \tag{13}$$

Further, an investment committed in a previous year should always be present in the consecutive years. This constraint is enforced by the following:
$$\forall l^y \in \Omega, \qquad n_l^y \geq n_l^{y-1} \tag{14}$$

Here, $n_l^y \geq 0$ and integer $\forall l^y \in \Omega$ and $l^y \neq k$; $(n_0^y + n_l^y - 1) \geq 0$ and integer for $l^y = k$, $k \neq 0$. $k = 0,1,\dots\omega_{cont}$, denotes the particular state of contingency, with $k = 0$ denoting the base case.

The primary motivation of this work is to demonstrate efficient techniques for ACTNEP with network contingencies and voltage stability constraint, which otherwise is computationally so demanding that it is almost impossible to solve the problem by conventional methods. As the RPP used for dynamic ACTNEP is considered same as what have been obtained in sequential ACTNEP, the constraints for the former do not include the usual constraints related to the additional reactive sources. However, these omitted constraints are completely considered when solving sequential planning [26].

## IV. SOLUTION TECHNIQUE

ACTNEP is a NP-hard problem with both integer and continuous variables which can be differentiated as:

$\forall\, y$, and $\forall\, k$, *State variables:* $V_{k_i}^y (\forall i \in N_{ld})$ and $\theta_{k_i}^y (\forall i \in N_{bus}, i \neq \text{slack})$; *Control variables:* $P_{Gn_{k_i}}^y (\forall i \in N_{pvbus}\; i \neq \text{slack})$, $V_{k_i}^y (\forall i \in N_{pvbus})$ and $n_l^y (\forall l^y \in \Omega)$; and *Fixed variables:* $P_{Dm_i}^y (\forall i \in N_{ld})$, $Q_{Dm_i}^y (\forall i \in N_{ld})$, $q_{rc_i}^y (\forall i \in N_{ld})$, and $\theta_{k_i}^y (i = \text{slack})$.

Solution of such a problem while considering all the variables as a single set is computationally intensive. However, computational complexity can be substantially reduced by suitable truncation of these variable sets and their successive solution. In this paper, it has been divided into two parts: a) investment variable part and b) operational variable part.

Line additions in a power corridor ($n_l^y$), which determine the investments and network topology, are obtained by MABC, while the estimation of power generations and voltage magnitudes at generator buses (operational variables) so as to satisfy network constraints, are performed by solving OPF (by in-built solvers in MATLAB). Through MABC, objective function (1) is minimized along with satisfaction of constraints (13) – (14). For a particular network topology fixed by MABC, OPF is solved for each network contingency to satisfy the remaining network and line flow constraints. Originally, power flow equations are non-linear in nature and to obtain adequate accuracy of planning, in base case topology, non-linear OPF is solved. However, repeated solutions of non-linear OPFs for each network contingency results in huge computational burden.

Therefore, to reduce the overall problem complexity, linearized OPF is solved in contingency cases. Such mixed form of solution methodology helps in obtaining a proper balance between the computational burden involved, and the accuracy of expansion planning.

Linearization of the network constraints are performed by assuming small angle difference between two connected nodes, and small deviation of voltage magnitudes from base values. By such assumption, for any two connected buses $i$ and $j$, it can be approximated that, $\sin\theta_{ij} \approx \theta_{ij}$ and $\cos\theta_{ij} \approx 1$. Here, $\theta_{ij} = \theta_i - \theta_j$. Substitution of these values in the evaluation of non-linear network constraints (3) – (4) and (9) – (12) and neglecting the higher order terms in Taylor's series expansion reduce them to linear constraints [29], which are solved effectively by OPF solver. At the end of OPF solution by the MATLAB solvers, fitness function values are returned to MABC, required for convergence to the optimal solution.

## V. Proposed Methodology

Solution of ACTNEP problems become exponentially complex when network contingencies considered. Further, compared to single-year static situation, when a multi-year DTNEP is considered, computational burden increases to a level, where rigorous, brute force solution methodologies cannot even be considered for use. As a result, solution to such problems for moderate to large systems require intelligent strategies that can efficiently obtain a good-quality solution.

In this paper, we propose a four-stage algorithm to quickly reach an acceptable solution for the N-1 security constrained, multi-year ACTNEP problems by application of several general intelligent strategies developed from the base case ACTNEP and contingency constrained DCTNEP results:

***Stage 1:*** *Solve base case DCTNEP.*
This stage is the easiest to solve and requires minimum computational burden, although it provides a logical starting point for the next stage of solving base case ACTNEP.

***Stage 2:*** *Solve base case ACTNEP.*
Starting from the results of previous stage [26], base case ACTNEP provides vital clues about the effective search space for contingency constrained ACTNEP.

***Stage 3:*** *Solve contingency constrained DCTNEP.*
Planning obtained from this stage provides a good starting point and viable estimation of the upper cost bound for contingency constrained ACTNEP.

***Stage 4:*** *Solve contingency constrained ACTNEP.*
Solution of this stage requires the maximum computational burden. Several strategies are formed from the results of the previous stages. These are applied with MABC to efficiently solve contingency constrained ACTNEP.

Results of the first three stages are much easier to obtain and provide valuable information for solving the fourth and final stage of N-1 security constrained ACTNEP. For Stage 4, the following strategies are proposed which drastically reduce the overall computational burden in solving security constrained AC dynamic TNEP:

### A. Estimate the set of power corridors in which the final solution will always be present

Computational burden in any optimization algorithm is directly proportional to its search space. A small search space reduces the computation burden for finding the optimum. In TNEP, all available power corridors of a system represent the search space. However, final contingency constrained ACTNEP solution shows new lines in only a few of all available power corridors. An estimation of these corridors with a very high possibility of having new lines in the final solution confines the search within this set and provides substantial reduction in computational burden. This violation set ($PC_{viol}$) can be obtained from the security analysis on base case ACTNEP solution. It provides all the power corridors where there are line power flow limit violations. Set of these power corridors ($PC_{viol} \in \Omega$) which is far lesser in size than the original set $\Omega$, provide a viable search space for the metaheuristic.

### B. Find the set of power corridors which will definitely have new lines in the final solution

Computational burden can be further reduced if a set of power corridors is precisely estimated which is sure to have new lines in the final security constrained planning. Such an estimation of fixed set ($PC_{fix}$) of power corridors allows MABC to always direct its search with in this set of power corridors which helps in faster arrival at the final solution. This set can be obtained by finding the common corridors present in set $PC_{viol}$ and in contingency constrained DCTNEP results ($DC_{cont}$).

Therefore, $PC_{fix} = PC_{viol} \cap DC_{cont}$.

### C. Reduce the number of times AC OPF is solved

The most time-consuming block in security constrained ACTNEP is the block that solves AC OPF. For each combination string generated by MABC, fitness function needs to be evaluated which involves solution of AC OPF. Computational burden can effectively be reduced by reducing the required number OPF solutions, as follows:

*1) Restrict the Number of Power Corridors Within a Specific Bound:*
It has been observed from solving ACTNEP for various systems that, in the final solution, the number of power corridors having new lines is almost 90% of the number present in corresponding DCTNEP results. ACTNEP will certainly have some more corridors than DCTNEP. In order to generalize the technique for use with both static and dynamic TNEP, the number of power corridors in solving ACTNEP are bounded within 90-130% of the number obtained in DCTNEP. Only when the number of power corridors with new lines in a combination string of MABC falls within this range, AC OPF is solved. In other cases, a suitable penalty is added to the objective function in order to discard the combination.

*2) Check the Worthiness of a Combination:*

In the initial phases of solution, most of the combinations generated by a metaheuristic are infeasible, which are gradually removed from the solution process by evaluating their fitness functions through solving OPFs. Hence, this makes the algorithm extremely inefficient as most of the time is spent in evaluating infeasible combinations. However, if only a combination deemed worthy of having a feasible solution is evaluated by solving AC OPF, the number of OPF solutions over the entire solution process reduces drastically. Worthiness of a combination is determined on the basis of its cost, and, OPF is solved, only if its cost is below a specific upper limit, ($U_{lim}$). For other cases, an appropriate penalty is added.

This limit is adaptively set as per the progress of the algorithm. Initially, it is set as twice the cost of new lines of security constrained DCTNEP. As the algorithm progresses, if a feasible combination with a lower investment cost is obtained, this cost is set as $U_{lim}$. Such a relaxed setting is used to allow MABC with sufficient flexibility of search to reach the final solution. Too tight a criterion to reject a combination may result in very constricted search space and may lead to trapping of the algorithm at a local optimum.

*3) Continue to Solve OPFs for Different Network Contingencies Only if Feasible Results are obtained for Base case and all Previous Contingencies:*

Final objective of security constrained TNEP is to obtain a planning which is feasible for every network configuration—base case and all network contingencies. For a combination produced by a metaheuristic, if the base case TNEP is not feasible, it is obvious that the contingency cases will also be infeasible. Further, once an infeasibility at any network contingency is obtained, remaining contingencies are not checked for feasibility, as it will eventually produce an infeasible final result. Suitable penalties are added to remove these infeasible combinations from the solution process. Therefore, computational burden in solving security constrained ACTNEP is effectively reduced by avoiding unnecessary OPF solutions.

*4) AC OPF is solved only for the years which Experience a change in the Base Topology:*

In the dynamic planning process, for a combination string generated, instead of solving the OPF block for all the years concerned, by this strategy, it is solved only for those years where there is a change in base network topology. Inclusion of this action is quite logical and produces a substantial reduction in the overall computational burden.

*D. Additional Reactive Sources are set at Values Obtained by solving Sequential ACTNEP*

Multi-year sequential ACTNEP involves sequential solving of static ACTNEP for each year concerned, with the planning at the end of a year becoming the base network for the next year. It is relatively much simpler to solve and results are obtained quite fast compared to DTNEP due to successive planning for every year. Also, sequential planning is shortsighted as it does not take into account future network conditions, and final investment cost obtained over the planning horizon is invariably higher than that obtained by dynamic planning. However, the results obtained from such a planning provides a good, sub-optimal starting point and upper bound for dynamic planning. It has been found by several trials of solving sequential and dynamic TNEP that, the values of the additional reactive sources obtained in both planning are very close. Therefore, to reduce the computational burden in solving multi-year AC DTNEP, the values of additional reactive sources are fixed to that obtained in sequential planning.

The proposed four-stage solution methodology although uses several intelligent strategies, it still retains sufficient flexibility to reach the final solution with drastic reduction of the computational burden. This property of the methodology will be evident in the next section, where detailed discussion on obtained results is done in comparison with rigorous method, which does not use any intelligent strategies. A detailed flow chart of the algorithm is depicted in Fig. 1.

## VI. RESULTS AND DISCUSSION

Applicability of the proposed methodology is demonstrated by solving security constrained multi-year dynamic ACTNEP for Garver 6 [22], IEEE 24 [22] and 118 [21] bus test systems. The systems considered provide an acceptable variation in size to demonstrate the suitability of the proposed methodology toward solving AC DTNEP from small to large systems. Sequential ACTNEP is also solved for these systems to demonstrate the benefits of DTNEP over sequential TNEP. As similar results are not available in present literature, comparison of results obtained by the proposed method with any other method is not possible. All generating units are considered to be completely dispatch-able. In base case, bus voltage magnitudes are constrained within $\pm 5\%$ of nominal values, whereas, for contingency cases, tolerance limit is $\pm 10\%$.

In each of the contingency cases, $pv$ bus voltage magnitudes and generations are modified so as to reduce line overloads, in accordance with actual practice. Simulations of this work are performed with MATLAB R2015b on a desktop computer with 16 GB RAM, having Intel (R) Core(TM) i5-4590 CPU processor @ 3.30 GHz. 50 trials for each system is performed and the best results are provided for comparison. In the solution procedure, stage 4 takes the maximum amount of time and compared to this, time required by the previous stages is considered negligible. Like any other metaheuristic, MABC also requires careful tuning of its parameters for optimum efficiency. The parameters are tuned according to the criterion in [26], with values for multi-year DTNEP provided in Table I. Detailed description of the methodology used for tuning the demands are considered to be 760 MW and 152 MVAR respectively for the first year. Dynamic TNEP for the system is carried out considering a planning horizon of three years. System load demands and discount rates are considered in accordance with [22]. Generation limits are considered as per yearly load demands.

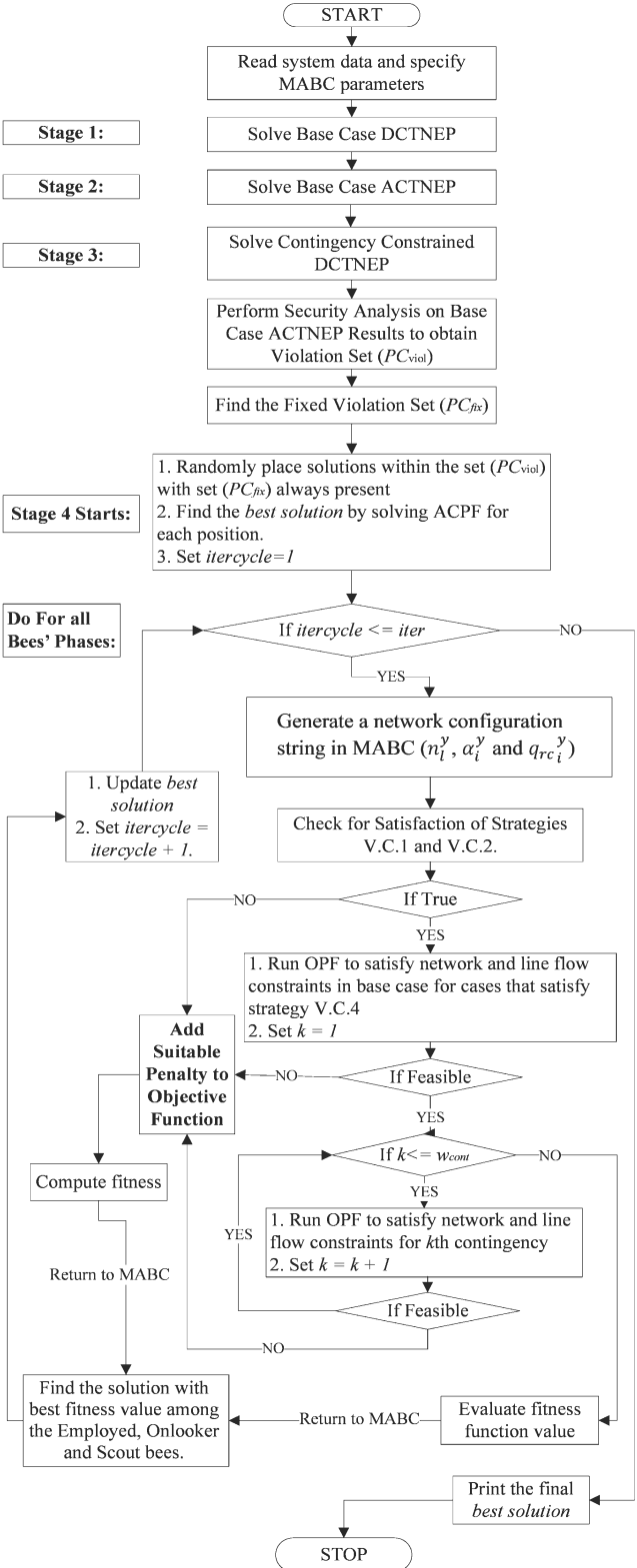

Fig. 1. Flow Chart of the Proposed Methodology

TABLE I
CONTROL PARAMETERS OF MABC

| Method | | $cs_N$ | $E_h$ | $lim$ | $iter$ | $w_g$ |
|---|---|---|---|---|---|---|
| Proposed | DCTNEP | 5 | 2 | 6 | 15 | 1.5 |
| | ACTNEP | 20 | 2 | 6 | 30 | 1.5 |
| Rigorous | ACTNEP | 20 | 2 | 6 | 30 | 1.5 |

A comparison of the planning results for base case DTNEP as obtained by the proposed method and single-stage rigorous method is shown in Table II. For solving base case TNEP, the strategies which are applicable to obtaining this solution are only applied. That is, strategies V.C.1, V.C.2 and V.D are only used. It can be observed from Table II that, both the rigorous and the proposed method obtains the same line addition costs as that obtained in [22]. However, time reduction obtained by the proposed method to obtain the solution is 98.87% compared to the rigorous method. This proves the applicability and efficiency of the proposed method.

TABLE II
DYNAMIC AC TNEP RESULTS OF GARVER 6 BUS SYSTEM FOR BASE CASE

| | Proposed Method | | | Rigorous Method | | |
|---|---|---|---|---|---|---|
| | Year 1 | Year 2 | Year 3 | Year 1 | Year 2 | Year 3 |
| New lines Constructed | $n_{1-5}=1$ $n_{2-3}=1$ $n_{2-6}=2$ $n_{3-5}=2$ $n_{4-6}=2$ | | $n_{2-6}=1$ $n_{3-5}=1$ | $n_{1-5}=1$ $n_{2-3}=1$ $n_{2-6}=2$ $n_{3-5}=2$ $n_{4-6}=2$ | | $n_{2-6}=1$ $n_{3-5}=1$ |
| No. of New Lines | 8 | 0 | 2 | 8 | 0 | 2 |
| $v_{lc}^y$ (x $10^3$ US\$) | 200 | 0 | 50 | 200 | 0 | 50 |
| $v_d$ (x $10^3$ US\$) | 223.900 | | | 223.900 | | |
| $L^y$ | 0.2713 | 0.3217 | 0.3608 | 0.2869 | 0.3214 | 0.3592 |
| $tp$ | 136.46 secs | | | 3.38 hrs | | |
| % Reduction in Computational Burden by Proposed Method | 98.87 | | | | | |

Planning results for security constrained DTNEP obtained by the proposed and single-stage rigorous methods are shown in Table III. It can be observed from the table that, similar to the previous case, both methods obtain similar line addition costs. However, reduction in computational burden obtained by the proposed method is 97.81%. This proves the effectiveness of the proposed method to reach the final solution with a drastic reduction in computational burden. Also, base case L-index values for all the years are observed to be well within the limit of 0.4.

*B. Parameter Tuning for the proposed Methodology*

In this section a detailed description on the procedure for an effective parameter tuning is provided. Performance of a metaheuristic algorithm is much dependent on the quality of its population pool. Also, to obtain a good solution, there has to be a sufficient level of variance within the population pool. More the variance, better is the chance of a metaheuristic algorithm parameters will be provided in a subsequent section.

*A. Multi-year Dynamic ACTNEP for Garver 6 Bus System*

This is a small system consisting of 6 buses with 15 power corridors. System data is obtained from [22] and a green-field expansion planning is considered. Total real and reactive power to obtain the global optimum solution. Otherwise, the solution is prone to get stuck in a local optimum. This essential feature required in the population of a metaheuristic algorithm is utilized to obtain an appropriate set of values of the different parameters of the proposed solution methodology. A particular value of the respective parameter that show maximum variance in the population pool compared to other values after a few iterations for several trials is considered a fair estimate for its optimal value.

TABLE III
SECURITY CONSTRAINED DYNAMIC ACTNEP RESULTS OBTAINED WITH THE PROPOSED AND RIGOROUS METHODS FOR GARVER 6 BUS SYSTEM

| Proposed Method | | | |
|---|---|---|---|
| | Year 1 | Year 2 | Year 3 |
| New lines Constructed | $n_{1-2}=1$; $n_{2-6}=3$; $n_{3-4}=1$ $n_{3-5}=4$; $n_{4-6}=2$ | $n_{2-3}=3$ $n_{4-6}=1$ | $n_{1-2}=1$ |
| No. of New Lines | 11 | 4 | 1 |
| $v_{lc}^y$ (x $10^3$ US$) | **329** | **90** | **40** |
| $v_d$ (x $10^3$ US$) | 413.7300 | | |
| $tp$ | 336.406 secs | | |
| $L^y$ | 0.1934 | 0.1688 | 0.1797 |
| Rigorous Method | | | |
| | Year 1 | Year 2 | Year 3 |
| New lines Constructed | $n_{1-2}=1$; $n_{2-6}=3$ $n_{3-4}=1$ $n_{3-5}=4$ $n_{4-6}=2$ | $n_{2-3}=3$ $n_{4-6}=1$ | $n_{1-2}=1$ |
| No. of New Lines | 11 | 4 | 1 |
| $v_{lc}^y$ (x $10^3$ US$) | **329** | **90** | **40** |
| $v_d$ (x $10^3$ US$) | 413.7300 | | |
| $tp$ | 4.27 hrs | | |
| $L^y$ | 0.1805 | 0.1748 | 0.1712 |
| Reduction in Computational Burden by Proposed Method | 97.81% | | |

TABLE IV
EFFECT OF $E_h$ ON SECURITY-CONSTRAINED DTNEP RESULTS FOR GARVER 6 BUS SYSTEM (WITH 5 TRIALS AND $lim=6$)

| $E_h$ | | 1 | **2** | 3 | 4 | 5 | 6 |
|---|---|---|---|---|---|---|---|
| Variance of population pool | 1st trial | 2.7901 | **5.0989** | 5.6138 | 1.6218 | 5.7966 | 3.6890 |
| | 2nd trial | 3.2992 | **4.4356** | 2.8771 | 1.3291 | 8.1304 | 4.9047 |
| | 3rd trial | 1.6127 | **15.3464** | 14.6887 | 5.2093 | 4.3461 | 6.0627 |
| | 4th trial | 5.0095 | **5.8443** | 2.7930 | 2.2491 | 4.3506 | 1.5330 |
| | 5th trial | 6.0395 | **17.2202** | 12.9559 | 5.1182 | 3.4245 | 3.8001 |
| Minimum cost (x $10^3$ US$) | | 588.350 | **498.900** | 444.450 | 588.350 | 550.884 | 583.480 |
| Maximum cost (x $10^3$ US$ | | 727.320 | **659.450** | 717.144 | 713.368 | 723.350 | 727.740 |
| Mean cost (x $10^3$ US$) | | 662.780 | **596.916** | 570.510 | 677.980 | 665.930 | 691.336 |
| Standard Deviation (x $10^3$ US$) | | 44.498 | **57.704** | 103.010 | 46.231 | 61.150 | 54.210 |

TABLE V
EFFECT OF $lim$ ON SECURITY-CONSTRAINED DTNEP RESULTS FOR GARVER 6 BUS SYSTEM (WITH 5 TRIALS, AND $E_h=2$)

| $lim$ | | 3 | 5 | **6** | 10 | 15 | 20 |
|---|---|---|---|---|---|---|---|
| Variance of population pool | 1st trial | 2.4512 | 3.6583 | **5.0989** | 6.7881 | 3.5789 | 5.6446 |
| | 2nd trial | 3.4404 | 4.7171 | **4.4356** | 1.1205 | 1.3062 | 4.1547 |
| | 3rd trial | 3.9021 | 2.9659 | **15.3464** | 8.7556 | 9.5734 | 6.5556 |
| | 4th trial | 5.7037 | 3.7013 | **5.8443** | 3.0036 | 5.5109 | 5.0880 |
| | 5th trial | 9.4456 | 3.0539 | **17.2202** | 4.6634 | 3.1733 | 12.2611 |
| Minimum cost (x $10^3$ US$) | | 564.320 | 562.450 | **498.900** | 605.984 | 588.350 | 680.580 |
| Maximum cost (x $10^3$ US$ | | 727.030 | 686.320 | **659.450** | 726.848 | 726.174 | 726.450 |
| Mean cost (x $10^3$ US$) | | 666.526 | 612.814 | **596.916** | 651.606 | 668.100 | 707.480 |
| Standard Deviation (x $10^3$ US$) | | 70.598 | 41.279 | **57.704** | 44.197 | 61.076 | 18.471 |

In the proposed solution methodology, there are two sets of parameters that need to be properly tuned: parameters related to MABC and parameters related to the intelligent strategies used.

For the MABC parameters, in case of $w_g$, its value is considered as 1.5 as per [30]. In [31], it is stated that the performance of ABC is not strongly dependent on colony size $cs_N$. As MABC is developed around the original ABC, it also shows a similar behavior and the value of $cs_N$ is kept at a conservative 20 for ACTNEP. An even lower value of 5 in case of DCTNEP is considered as it is much easy to solve compared to ACTNEP. Also, it has been found that, the final solution is obtained within about 30 iterations in all the test cases. Therefore, value of $iter$ is set at 30 for ACTNEP. For DCTNEP, again a lower value of $iter$ is considered due to the reasons stated previously. Thus, the MABC parameters required to be properly tuned are only $E_h$ and $lim$. For tuning of these parameters, rigorous solution of security constrained Dynamic ACTNEP for Garver 6 bus system is performed. Here, one variable is kept fixed and the other is gradually changed to find its best value. The results obtained by such changes is shown in Tables IV and V. It can be observed from these tables that, highest amount of variance in the population is obtained only with $E_h$ value of 2 and $lim$ value of 6. Therefore, these values are considered for solving sequential and DTNEP for the various test systems by the proposed method.

After the parameters for MABC are tuned, it is required to tune the parameters related to the intelligent strategies used. For the intelligent strategies, the most important parameter is the bounding of the number of power corridors as a percentage of the number in DCTNEP results. Therefore, this bound needs to be properly tuned. Same procedure as used in the previous tuning procedure is also used for this case. Security constrained DTNEP for 6 bus system is solved by the proposed method for this tuning process. The particular bound which provides the highest variance in the population pool is considered as the best bound. Results for this variation are provided in Table VI. As the corridor bound of 90-130% provides the highest variance in the population pool, this bound is used for solution of sequential and DTNEP of the test systems.

### C. Multi-year Dynamic ACTNEP for IEEE 24 Bus System

This 24-bus system has a real and reactive power demand of 8550 MW and 1740 MVAR respectively in first year. There are 41 power corridors with each corridor having the ability to accommodate a maximum of 3 new lines. System data and installation costs have been obtained from [22]. A three-year planning horizon is considered, with load increment and discount costs similar to the previous case. Due to extreme computational burden experienced in solving multi-year ACTNEP by rigorous method, solution is only obtained by the proposed method.

TABLE VI
EFFECT OF PERCENTAGE MATCHING OF CORRIDORS ON SECURITY-CONSTRAINED DTNEP RESULTS FOR GARVER 6 BUS SYSTEM (WITH 5 TRIALS, $E_h = 2$ AND $lim = 6$)

| Percentage matching of corridors | | 70-200% | 70-150% | 80-150% | 90-150% | **90-130%** | 90-200% |
|---|---|---|---|---|---|---|---|
| Variance of population pool | 1st trial | 5.4066 | 7.7541 | 21.8641 | 12.2707 | **10.3884** | 15.2393 |
| | 2nd trial | 9.4667 | 14.5107 | 4.7658 | 4.9605 | **7.0348** | 1.9990 |
| | 3rd trial | 8.3063 | 21.8774 | 15.0616 | 1.6646 | **31.6930** | 12.5272 |
| | 4th trial | 8.0107 | 13.5674 | 9.1951 | 5.8764 | **34.7697** | 2.6933 |
| | 5th trial | 9.9602 | 7.4006 | 4.5130 | 10.0106 | **5.8265** | 10.9062 |
| Minimum cost (x $10^3$ US$) | | 573.160 | 588.990 | 529.700 | 467.870 | **680.410** | 643.610 |
| Maximum cost (x $10^3$ US$) | | 685.160 | 721.840 | 719.610 | 722.740 | **715.350** | 703.594 |
| Mean cost (x $10^3$ US$) | | 637.830 | 657.710 | 662.920 | 592.870 | **695.900** | 685.650 |
| Standard Deviation (x $10^3$ US$) | | 45.267 | 45.662 | 68.673 | 86.713 | **11.742** | 59.480 |

Detailed solution procedure of obtaining the results for the first-year security constrained sequential ACTNEP by the proposed methodology can be described as follows:

*1) Stage 1: Base case DCTNEP*

Solution of base case DCTNEP is obtained extremely fast (in approx. 9.75 secs) with a planning cost of 78 x$10^6$ US$. New lines are obtained in corridors 6-10, and 13-14.

*2) Stage 2: Base case ACTNEP*

Starting from stage 1, base case ACTNEP results are obtained according to the procedure described in [27]. The final planning cost obtained is 98 x$10^6$ US$ with new lines in power corridors 6-10, 7-8 and 11-13, i.e. in corridor numbers 13, 14 and 21. Solution time is only about 100 secs.

*3) Stage 3: Contingency Constrained DCTNEP*

Solution of this stage provides a planning cost of 376 x$10^6$ US$ with new lines in corridor numbers 3, 8, 10, 13, 14, 19, 21, 28 and, 40. So, $DC_{cont} = \{3, 8, 10, 13, 14, 19, 21, 28, 40\}$, i.e. 9 corridors have new lines, solution time of approx. 135 secs.

*4) Stage 4: Contingency Constrained ACTNEP*

This stage starts by estimation of the sets $All_{viol}$ and $PC_{fix}$. Set $PC_{viol}$ is obtained by performing security analysis on stage 2 results. Represented by corridor numbers, this is,
$PC_{viol} = \{1, 2, 3, 5, 6, 8, 9, 10, 11, 13, 14, 15, 18, 19, 20, 21,$
$22, 23, 24, 26, 27, 28, 29, 30, 31, 32, 33, 35, 36, 37, 40, 41\}$
Therefore, $PC_{fix} = \{3, 8, 10, 13, 14, 19, 21, 28, 40\}$.

Application of strategy A. of Section V confines the search within 32 of 41 available corridors, thereby providing a reduction of 21.95% in the search space. Further, strategy V.B. forces MABC to always include set $PC_{fix}$ in its search. Next, V.C. reduces the required number of OPF solutions by the application of three strategies V.C.1, V.C.2 and V.C.3. With the application of V.C.1, OPF is solved only if the number of corridors with new lines in a combination is within 90-130% of 9, i.e. within 8 and 12 corridors. Further, by V.C.2, OPF is solved only when cost of new lines in a combination is less than $U_{lim}$. It can be mathematically represented as:

Solve AC OPF if, $\forall b \in$ all combinations, $y = 1$,

$$v_{lc_b}^y < U_{lim} \quad \& \quad 8 \leq \sum_{n_{l\,b}^y > 0} l_b^y \leq 12 \quad (20)$$

By V.C.3, OPF is not required to be solved for a combination for remaining network configurations once an infeasibility is obtained. Table VII provides the results for both sequential and dynamic ACTNEP. The final result also shows that for first-year sequential TNEP, new lines are confined within 10 power corridors of $PC_{viol}$ as provided by strategy V.C. It can be further observed from Table VII that by solving dynamic TNEP, it is possible to obtain an investment cost, which is 5.48% lower than that obtained by sequential TNEP. This translates to savings of 44.96 x $10^6$ US$, which is a substantial amount, while increase in computational burden in solving dynamic TNEP over sequential is around 38.71%. Base case L-index values are limited within 0.4 as per the system stability constraint used. Further restriction of their values to even lower limits will result in greater system stability at an increased investment cost and vice-versa.

TABLE VII
Security Constrained Multi-year ACTNEP results obtained with the proposed method for IEEE 24 bus system

| **Dynamic Planning** | | | |
|---|---|---|---|
| | Year 1 | Year 2 | Year 3 |
| New lines Constructed | $n_{1-5} = 1$; $n_{2-4} = 1$; $n_{3-9} = 1$ $n_{4-9} = 1$; $n_{6-10} = 2$; $n_{7-8} = 3$ $n_{10-11} = 1$; $n_{11-13} = 2$ $n_{14-16} = 1$; $n_{20-23} = 1$ | $n_{3-9} = 1$; $n_{6-10} = 1$ $n_{9-11} = 1$; $n_{14-16} = 1$ $n_{16-17} = 1$; $n_{19-20} = 1$ $n_{21-22} = 1$ | $n_{10-11} = 1$ $n_{15-21} = 1$ $n_{20-23} = 1$ |
| No. of New Lines | **14** | **7** | **3** |
| $v_{lc}^y$ (x $10^6$ US$) | **459** | **336** | **148** |
| $v_d$ (x $10^6$ US$) | 774.6880 | | |
| $tp$ | 4788.719 secs | | |
| $L^y$ | 0.3478 | 0.3741 | 0.3968 |
| **Sequential Planning** | | | |
| | Year 1 | Year 2 | Year 3 |
| New lines Constructed | $n_{1-5} = 1$; $n_{3-9} = 1$ $n_{4-9} = 1$; $n_{6-10} = 2$ $n_{7-8} = 3$; $n_{10-11} = 1$ $n_{11-13} = 1$; $n_{14-16} = 1$ $n_{14-23} = 1$; $n_{20-23} = 1$ | $n_{3-24} = 1$; $n_{6-10} = 1$ $n_{11-13} = 1$; $n_{15-24} = 1$ $n_{16-17} = 1$; $n_{17-22} = 1$ | $n_{1-2} = 1$ $n_{1-5} = 1$ $n_{9-11} = 1$ $n_{10-11} = 1$ $n_{15-21} = 1$ |
| No. of New Lines | 13 | 6 | 5 |
| $v_{lc}^y$ (x $10^6$ US$) | 446 | 386 | 193 |
| $v_d$ (x $10^6$ US$) | 819.6480 | | |
| $tp$ | 3452.385 secs | | |
| $L^y$ | 0.3710 | 0.3447 | 0.3264 |
| **Reduction in Overall Cost by Dynamic TNEP Compared to Sequential TNEP** | | | **5.46%** |

## D. Multi-year Dynamic ACTNEP for IEEE 118 Bus System

This large system consists of 118 buses and 179 physical power corridors [31]. Out of 179, seven power corridors consist of two different sets of lines, thereby in actual totalling to 186 power corridors to be considered. It consists of 54 generators and 91 loads. Total real and reactive power demands are respectively 3733.07 MW and 1442.98 MVAR. Line capacities are reduced to 60% of original to create network congestion. Line construction costs are estimated as in [20].

For a large system as this, computational burden in solving ACTNEP by rigorous method is prohibitively large. Therefore, multi-year ACTNEP is only solved by the proposed method. Results obtained are provided in Table VIII. It can be observed that, the proposed method obtained both sequential and DTNEP results within a manageable time frame. Savings of 7.3018 x $10^6$ US\$ in investment cost is obtained by solving DTNEP over sequential TNEP, with high system stability in both the cases.

TABLE VIII
Security Constrained Multi-year ACTNEP results obtained with the proposed method for IEEE 118 bus system

| | Dynamic Planning | | |
|---|---|---|---|
| | Year 1 | Year 2 | Year 3 |
| New lines Constructed | $n_{8-5}$ = 1; $n_{23-32}$ = 1 $n_{38-37}$ = 1; $n_{38-65}$ = 1 $n_{64-65}$ = 1; $n_{77-78}$ = 1 $n_{94-100}$ = 1 | $n_{65-68}$ = 1 $n_{94-95}$ = 1 $n_{94-100}$ = 1 $n_{99-100}$ = 1 | $n_{2-12}$ = 1; $n_{17-18}$ = 1 $n_{30-17}$ = 1; $n_{26-30}$ = 1 $n_{34-37}$ = 1 |
| No. of New Lines | 7 | 4 | 5 |
| $v_{lc}^y$ (x $10^6$ US\$) | 66.4665 | 36.4860 | 44.7570 |
| $v_d$ (x $10^6$ US\$) | 114.4586 | | |
| $tp$ | 4.02 hrs | | |
| $L^y$ | 0.0676 | 0.0827 | 0.0947 |
| | Sequential Planning | | |
| | Year 1 | Year 2 | Year 3 |
| New lines Constructed | $n_{8-5}$ = 1; $n_{23-32}$ = 1 $n_{38-65}$ = 1; $n_{64-65}$ = 1 $n_{77-78}$ = 1 | $n_{8-30}$ = 1; $n_{34-37}$ = 1 $n_{38-37}$ = 1; $n_{65-68}$ = 1 $n_{80-99}$ = 1; $n_{94-100}$ = 1 | $n_{5-11}$ = 1; $n_{2-12}$ = 1 $n_{17-18}$ = 1; $n_{26-30}$ = 1 $n_{94-95}$ = 1 |
| No. of New Lines | 5 | 6 | 5 |
| $v_{lc}^y$ (x $10^6$ US\$) | 49.2765 | 62.8770 | 55.7460 |
| $v_d$ (x $10^6$ US\$) | 121.7604 | | |
| $tp$ | 2.76 hrs | | |
| $L^y$ | 0.0678 | 0.0828 | 0.0945 |
| **Reduction in Overall Cost by Dynamic TNEP Compared to Sequential TNEP** | **6.00%** | | |

## VII. CONCLUSIONS

This paper proposes a four-stage solution methodology to efficiently solve non-linear, multi-year security constrained ACTNEP problems, which is not attempted in the past. Final planning results are obtained extremely quickly from the results of AC base case and DC security constrained planning with the use of several novel and intelligent strategies. Proper parameter tuning is done to obtain good quality final solutions. The strategies lead to a significant narrowing down of the overall search space for finding optimum in the final stage. Although, there is a reduction in search space in the final AC stage, it does not hamper the algorithm convergence process as the previous stages consider the entire space for obtaining respective results. It is also evident from the comparison with results obtained by traditional, rigorous method. However, application of the proposed methodology provides near 95% reduction in computational burden compared to rigorous method. The planning topologies also provide a high degree of voltage stability to the system as is evident from the low L-index values in all case studies. Further, Tables VII and VIII show that the results are obtained by the proposed method even for a large system within manageable time-frame and computational burden.

The strategies developed in this work are derived in accordance with the results obtained in previous stages of solution process. Hence, these are not system specific, and general enough to be used in any system, with any solution algorithm or metaheuristic. Therefore, by the use of these strategies, in future, solution of complex TNEP problems involving uncertainty, dynamic stability, etc. can be obtained with tremendous efficiency.